\documentstyle[12pt,epsf]{article} 

\def\lsim{\mathrel{\lower2.5pt\vbox{\lineskip=0pt\baselineskip=0pt 
           \hbox{$<$}\hbox{$\sim$}}}} 
\def\gsim{\mathrel{\lower2.5pt\vbox{\lineskip=0pt\baselineskip=0pt 
           \hbox{$>$}\hbox{$\sim$}}}}

\def\k{\kappa}
\def\p{\partial}

\def\d{\delta}

\def\L{\Lambda}

\def\h{h_{\mu\nu}}
\def\hp{\hat{\psi}}
\def\tp{\tilde{\psi}}
\begin{document} 
\begin{flushright}
DPNU-01-33\\ hep-th/0112224
\end{flushright}

\vspace{10mm}

\begin{center}
{\Large \bf 
 Newton's law in braneworlds with an infinite extra dimension}

\vspace{20mm}
 Masato ITO 
 \footnote{E-mail address: mito@eken.phys.nagoya-u.ac.jp}
\end{center}

\begin{center}
{
\it 
{}Department of Physics, Nagoya University, Nagoya, 
JAPAN 464-8602 
}
\end{center}

\vspace{25mm}

\begin{abstract}
 We study the behavior of the four$-$dimensional Newton's law in 
 warped braneworlds.
 The setup considered here is a $(3+n)$-brane embedded in $(5+n)$ 
 dimensions, where $n$ extra dimensions are compactified and a dimension
 is infinite.
 We show that the wave function of gravity is described in terms of
 the Bessel functions of $(2+n/2)$-order and that estimate the
 correction to Newton's law.
 In particular, the Newton's law for $n=1$ can be exactly obtained.
\end{abstract} 

\newpage 
%
%
 \section{Introduction}
 
 Recent ideas of extra dimensions have developed rapidly and its
 investigations have become a center of particle physics and cosmology.
 It is satisfactory to consider that our four-dimensional world is
 embedded in higher dimensional world.
 This picture which comes from sting/M-theory is so-called braneworld
 which assumes that Standard model fields are confined to a $3$-brane
 with four-dimensional spacetime.
 Motivated by the warped braneworld \cite{Rubakov:1983bz},
 in the framework of simple braneworld with two $3$-branes embedded in
 $AdS_{5}$, Randall and Sundrum proposed a new suggestion to the
 hierarchy problem owing to warped metric \cite{Randall:1999ee}.
 On the other hand, the warped geometry including a bulk scalar field
 has various types \cite{Ito:2001jk,Ito:2001gk}.
 Moreover interestingly the gravity in this warped geometry exhibits an
 interesting behavior.
 In the Randall-Sundrum model, graviton is localized on the positive
 tension $3$-brane, and usual four-dimensional Newton's law
 can be recovered at distance which is much larger than a radius of
 Anti-de Sitter space \cite{Randall:1999vf}.
 In particular, the four-dimensional effective Planck scale can be
 finite even if extra dimension is infinite, namely, non-compact 
 \cite{Rubakov:2001kp}.
 This implies that zero mode of gravity becomes a bound state and
 massive mode of one has continuous eigenvalue.
 In various model the localization of gravity is widely discussed
 \cite{Lykken:1999nb,Giddings:2000mu,Karch:2000ct,Csaki:2000fc}.

 In this paper, we study the four-dimensional Newton's
 law on a $(3+n)$-brane embedded in $(5+n)$-dimensional world with
 negative bulk cosmological constant.
 It is assumed that $n$ extra dimensions on the brane are
 compactified in same radius $R\sim M^{-1}_{\rm Pl}$, where $M_{\rm Pl}$
 is Planck scale, and that a perpendicular direction to brane
 is non-compact with $Z_{2}$ symmetry.
 This setup corresponds to an extension of the five-dimensional
 Randall-Sundrum model to $(5+n)$-dimensions.
 Taking account of four-dimensional gravitational fluctuation about the
 background metric, we show that wave function of gravitational field is
 expressed in terms of the Bessel functions of $(2+n/2)$-order.
 In order to investigate the behavior of gravity, we calculate the
 correction to the four-dimensional Newton's law. 
 We provide a possibility of searching for the remnant of
 compactification  even if the compactified radius with Planck length is
 invisible in our world.  

 In section 2 we describe a setup considered here and show the wave
 function of gravity.
 In section 3 we obtain the probability for existence of gravity on the
 brane and estimate the correction to four-dimensional Newton's law.
 In final section we describe a summary.
 
%
 \section{Setup}
 We consider the model of a single $(3+n)$-brane embedded 
 in the bulk $(5+n)$-dimensions, and the action is given by
  \begin{eqnarray}
  S=\int d^{4+n}x\;dy\sqrt{-G}
    \left(\;\frac{1}{2\k^{2}}{\cal R}
    -\L\;\right)-\int d^{4+n}x\;\sqrt{-g}\;V\,,\label{eqn1}
 \end{eqnarray}
 where ${\cal R}$ is the higher dimensional scalar curvature,
 $\L$ is the cosmological constant in the bulk and $V$ is brane
 tension, and $1/\k^{2}$ is the higher dimensional fundamental
 scale which has mass dimension $3+n$.
 Let $G$ be the metric in the bulk and $g$ be the metric induced on the
 brane, $g_{\mu\nu}=G_{\mu\nu}(y=0)$.
 Here the ansatz for $(5+n)$-dimensional metric is taken as follows 
 \begin{eqnarray}
  ds^{2}&=&a^{2}(y)\left(\eta_{\mu\nu}dx^{\mu}dx^{\nu}
        +\sum^{n}_{i=1}\;dz^{2}_{i}\right)+dy^{2}\nonumber\\
  &\equiv&G_{MN}dx^{M}dx^{N}\,,\label{eqn2}
 \end{eqnarray}
 where $\eta_{\mu\nu}=(-,+,+,+)$ and $a(y)$ is warp factor.
 It is assumed that $z$-directions are compactified in
 the same radius $R$ of Planck size and $y$-direction is an infinite
 extra dimension.
 This setup is as same as one given in Ref \cite{Dubovsky:2000av}
 which discussed the four-dimensional electrodynamics on the brane
 embedded in higher dimensions and the localization of $U(1)$ gauge
 field was investigated in detail.
 However we are interested in the shape of gravity in this model.
 This situation corresponds to a special case (isotropic brane tension)
 described in Ref \cite{Ito:2001fd}.
 Solving the Einstein equation of this model with negative bulk
 cosmological constant, we have 
 \begin{eqnarray}
  a(y)=e^{-|y|/L}\,,
  \hspace{0.5cm}\frac{1}{L}=\sqrt{\frac{-2\k^{2}\L}{(n+3)(n+4)}}\,,
 \label{eqn3}
 \end{eqnarray}
 where $a(y)$ respects the $Z_{2}$ symmetry $y\sim -y$.
 The jump condition with respect to the derivative of $a$ at $y=0$
 leads to the value of positive brane tension 
 \begin{eqnarray}
  V=\frac{2(n+3)}{\k^{2}L}\,.\label{eqn4}
 \end{eqnarray}
 As is evident from the above equations, the case of $n=0$ is completely
 consistent with the solution in the original Randall-Sundrum model.
 
 Integrating out the extra dimensions, squared Planck scale is identified
 with the coefficient of the four-dimensional scalar curvature
 and we obtain
 \begin{eqnarray}
  M^{2}_{\rm Pl}\sim 
  \frac{1}{\k^{2}}\int d^{n}z\int^{\infty}_{0}dy\; \left[a(y)\right]^{n+2}
  \sim \frac{R^{n}}{\k^{2}}L\,.\label{eqn5}
 \end{eqnarray}
 In order to study the behavior of four-dimensional gravity,
 the gravitational fluctuation around the background metric is given by
 \begin{eqnarray}
  G_{\mu\nu}=a^{2}(y)\eta_{\mu\nu}+h_{\mu\nu}(x,y,z)\,.\label{eqn6}
 \end{eqnarray}
 By imposing transverse-traceless gauge for these fluctuations, 
 {\it i.e.} $\p_{\mu}h^{\mu}_{\nu}=h^{\mu}_{\mu}=0$, the wave function
 of gravity is governed by the familiar eigenvalue problem of the 
 Schr$\ddot{\rm o}$dinger equation. 
 In the context of this situation we treat the fluctuation of
 four-dimensional part, the additional fluctuations not of the form in 
 transverse-traceless gauge are neglected.
 As for this point, we will describe it somewhere.

 We calculate the effective four-dimensional static gravitational
 potential between two masses placed on the positive tension brane at a
 distance $r$ each other.
 It is assumed that $r$ is much larger than
 compactified radius $R\sim M^{-1}_{\rm Pl}$ and
 that $L$ is set to the intermediate scale, $R\ll L$.
 Although the gravity propagating in the bulk has massive KK-modes along
 $n$ compactified $z$-directions, these dimensions are invisible at
 long distance under the above assumption.
 Namely, the dependence of $z$ in $\h$ is neglected here.
 Performing a separation of variables $x,y$ in wave function,
 $\h(x,y)=\h(x)\psi(y)$, and a change of variable
 $\psi(y)=a^{-n/2}(y)\;\hp(y)$ yields
 \begin{eqnarray}
  \left[\;\frac{d^{2}}{dy^{2}}+m^{2}e^{2|y|/L}
  -\frac{(n+4)^{2}}{4L^{2}}+\frac{n+4}{L}\d(y)\;\right]\hp(y)=0
  \label{eqn7}\,,
 \end{eqnarray}
 where $m^{2}$ is the four-dimensional mass.
 Moreover, following the change of variable given in Ref
 \cite{Randall:1999vf},
 $|u|=L(e^{|y|/L}-1)$ and $\hp(y)=\tp(u)e^{-|y|/2L}$,
 the above equation leads to the non-relativistic quantum mechanics
 problem as follows
 \begin{eqnarray}
  \left[\;-\frac{d^{2}}{du^{2}}+V(u)\;\right]\tp^{(n)}(u)=m^{2}\tp^{(n)}(u)
  \,,\label{eqn8}
 \end{eqnarray}
 where a superscript $n$ in wave function represents the number of
 extra dimensions in $(3+n)$-brane, a volcano potential given here is
 given by
 \begin{eqnarray}
  V(u)=\frac{(n+4)^{2}-1}{4(|u|+L)^{2}}-\frac{n+3}{L}\d(u)\,.
  \label{eqn9}
 \end{eqnarray}
 Here a potential form in the Schr$\ddot{\rm o}$dinger equation for $n=0$
 corresponds to a one in the original Randall-Sundrum model.
 An attractive potential of delta-function type causes the gravity
 to have a bound state.
 From Eq.(\ref{eqn8}), the zero mode wave function with $m^{2}=0$ can be
 normalizable, we have
 \begin{eqnarray}
  \hp_{0}(u)=\sqrt{\frac{n+2}{2}}\;L^{(n+2)/2}
  \left(|u|+L\right)^{-(n+3)/2}\label{eqn10}\,.
 \end{eqnarray}
 The wave function with continuous mode can be expressed
 in terms of superposition of the Bessel functions of $(2+n/2)$ order
 \begin{eqnarray}
  \hp_{m}(u)=\left(|u|+L\right)^{1/2}
  \left\{\;AJ_{2+n/2}\left(m(|u|+L)\right)
         +BY_{2+n/2}\left(m(|u|+L)\right)\;\right\}\label{eqn11}\,,
 \end{eqnarray}
 where $J_{\nu}$ and $Y_{\nu}$ is the Bessel functions of the first kind
 and ones of the second kind, respectively, $A$ and $B$ are
 constants to be determined below.
 The jump condition at $y=0$ leads to the relation between $A$ and $B$,
 so that $A=-BY_{1+n/2}(mL)/J_{1+n/2}(mL)$.
 Furthermore, the normalization factor in wave function $\tp_{m}$ can 
 be determined by the orthonormalization condition of Bessel functions.
 Thus we get 
 \begin{eqnarray}
  \lefteqn{\tp_{m}(u)=N_{m}\left[m\left(|u|+L\right)\right]^{1/2}}
  \nonumber\\
   &&\times\left[\;
  -Y_{1+n/2}(mL)J_{2+n/2}(m(|u|+L))
  +J_{1+n/2}(mL)Y_{2+n/2}(m(|u|+L))\;\right]\label{eqn12}\,, 
 \end{eqnarray}
 where
 \begin{eqnarray}
  N_{m}=\frac{1}
  {\sqrt{\left[J_{1+n/2}(mL)\right]^2+\left[Y_{1+n/2}(mL)\right]^2}}\,.
  \label{eqn13}
 \end{eqnarray}

 Obviously, it is fact that the behaviors of Bessel functions are
 quite different forms whether $n$ even or odd. 
 The Bessel functions of odd order are written in terms of
 sine or cosine functions, explicitly.
 Actually, it is important to study the form of $N_{m}$ since the
 correction to the Newton's law is investigated by integral of mode $m$.
 For $n$ even, the normalization factor $N_{m}$ is written
 by the asymptotic form of Bessel functions depending on the magnitude
 of argument.
 On the other hand, in the case that $n$ is odd, the exact form of
 $N_{m}$ can be described in terms of elementary functions
 \footnote[2]{
 The exact form of $N_{m}$ is determined by using 
 familiar relations between half-integer order Bessel
 functions of first kind and second kind as follows, \\
  $\displaystyle\left[ J_{3/2}(x)\right]^{2}+\left[Y_{3/2}(x)\right]^{2}=
  \frac{2}{\pi}\left(x^{-1}+x^{-3}\right)\,,$\\
  $\displaystyle\left[J_{5/2}(x)\right]^{2}+\left[Y_{5/2}(x)\right]^{2}=
  \frac{2}{\pi }\left(9x^{-3}+(3-x^{2})^{2}x^{-5}\right)\,,$\\
  $\displaystyle\left[J_{7/2}(x)\right]^{2}+\left[Y_{7/2}(x)\right]^{2}=
  \frac{2}{\pi }\left((15-x^{2})^{2}x^{-5}+(15-6x^{2})^{2}x^{-7}\right)\,.$
 }.

 In present model, in despite of neglecting the dependence of 
 compactified directions in four-dimensional fluctuation,
 it is found that the form of wave function depends on the number of
 compactified extra dimensions in the brane.

%
 \section{Correction to 4-d Newton's law}

 The correction to the four-dimensional Newton's law between two masses
 $M_{1}$ and $M_{2}$ is generated by the exchange of the massive modes,
 it is given by \cite{Randall:1999vf}
  \begin{eqnarray}
 U_{n}(r)=
  \frac{\k^{2}}{R^{n}}
  \int^{\infty}_{0}dm\;M_{1}M_{2}
  \frac{e^{-mr}}{r}\left|\tp^{(n)}_{m}(0)\right|^{2}\,,\label{eqn14}
 \end{eqnarray}
 where the probability for existence of gravity with continuous
 mode on the brane at $y=0$ is
 \begin{eqnarray}
  \left|\tp^{(n)}_{m}(0)\right|^{2}=
  \frac{4}{\pi^{2}mL}\frac{1}
  {\left[J_{1+n/2}(mL)\right]^2+\left[Y_{1+n/2}(mL)\right]^2}\,.
  \label{eqn15}
 \end{eqnarray}
 Here we used Eqs. (\ref{eqn12}), (\ref{eqn13}) and Lommel's formula 
 $J_{\nu+1}(x)Y_{\nu}-J_{\nu}Y_{\nu+1}(x)=2/(\pi x)$. 
 The coefficient of the integral in Eq.(\ref{eqn14}) is
 the effective five-dimensional gravitational constant because the
 gravity is essentially five-dimensions with an infinite extra dimension
 at long distance $r\gg R$.
 The integral of Eq.(\ref{eqn14}) depends on the magnitude of
 argument in the Bessel functions of Eq.(\ref{eqn15}).

 For $n$ even (this implies that Bessel functions of integer
 order), at $mL\ll 1$, the Bessel function of second kind is obviously
 dominant in denominator of Eq.(\ref{eqn15}).
 On the other hand, at $mL\gg 1$, the asymptotic behaviors of Bessel
 functions are written in terms of sine or cosine functions,
 $\sqrt{z}J_{\nu}(z)\sim \sqrt{2/\pi}\cos z$ and 
 $\sqrt{z}Y_{\nu}(z)\sim \sqrt{2/\pi}\sin z$ for $z\gg 1$.
 Namely, Eq.(\ref{eqn15}) becomes constant.
 Thus we obtain that $|\tp^{(n)}_{m}(0)|^{2}\sim (mL)^{1+n}$ for
 $m\ll L^{-1}$ and $|\tp^{(n)}_{m}(0)|^{2}\sim 2/\pi$ for 
 $m\gg L^{-1}$.
 Consequently,  it is necessary to divide this integral into two
 regions, $mL\ll 1$ and $mL\gg 1$, and Eq.(\ref{eqn14}) is expressed as
 \begin{eqnarray}
   U_{n}(r)=
  G_{N}L\frac{M_{1}M_{2}}{r}
  \left[\;
  \int^{L^{-1}}_{0}dm\;(mL)^{1+n}e^{-mr} 
 +\int^{\infty}_{L^{-1}}dm\;\frac{2}{\pi}e^{-mr} \;\right]\;\,,
  \label{eqn16}
 \end{eqnarray}
 Here we used $G_{N}\sim M^{-2}_{\rm Pl}\sim \k^{2}/(R^{n}L)$,
 where $G_{N}$ is the four-dimensional Newton constant.
 The first term in Eq.(\ref{eqn16}) is the contribution of the light
 mode $(m\ll L^{-1})$ and the second term is the contribution of the
 heavy mode $(L^{-1}\ll m\ll R^{-1}\sim M_{\rm Pl})$.
 \begin{eqnarray}
  U_{n}(r)=
  G_{N}L\frac{M_{1}M_{2}}{r}
  \left[\;L^{1+n}
  \left(-\frac{\p}{\p r}\right)^{1+n}
  \frac{1-e^{-r/L}}{r}
 +\frac{2}{\pi}\frac{e^{-r/L}}{r} \;\right]\;\,,\label{eqn17}
 \end{eqnarray}
 At distance $r\gg L$, the first term in Eq.(\ref{eqn17}) is dominant.
 From Eqs.(\ref{eqn10}) and (\ref{eqn14}),
 the usual four-dimensional Newton's law can be recovered via the
 contribution of zero mode $\hp_{0}(0)\sim 1/\sqrt{L}$.
 This means that the four-dimensional effective Planck scale is finite,
 as indicated in Eq.(\ref{eqn5}).
 Consequently, the four-dimensional Newton's law $V_{n}(r)$ by adding
 the contribution of zero mode is
 \begin{eqnarray}
  V_{n}(r)\sim G_{N}\frac{M_{1}M_{2}}{r}
  \left(\;1+C\left(\frac{L}{r}\right)^{2+n}\;\right)\,,\label{eqn18}
 \end{eqnarray}
 where $C$ is a numerical constant.
 The second term in the bracket is the contribution of short distance
 correction to Newton's law, which is consistent with the original
 Randall-Sundrum model.
 Moreover, at distance $r\ll L$,
 the second term in Eq.(\ref{eqn17}) is dominant and we have
 \begin{eqnarray}
 V_{n}(r)\sim G_{N}L\frac{M_{1}M_{2}}{r^{2}}\,,\label{eqn19}
 \end{eqnarray}
 where corresponds to the five-dimensional Newton's law.
 In this situation, we cannot see the four-dimensional world.

 For $n$ odd, as shown in footnote $\dagger$, we can obtain the exact
 form of normalization factor $N_{m}$ in wave function.
 Consequently, the probability for existing gravity on the
 brane is given by
 \begin{eqnarray}
 \left|\tp^{(1)}_{m}(0)\right|^{2}
        &=&\frac{2}{\pi}\frac{(mL)^{2}}{(mL)^{2}+1}\nonumber\\
 \left|\tp^{(3)}_{m}(0)\right|^{2}
        &=&\frac{2}{\pi}\frac{(mL)^{4}}{(mL)^{4}+3(mL)^{2}+9}\nonumber\\
 \left|\tp^{(5)}_{m}(0)\right|^{2}
        &=&\frac{2}{\pi}\frac{(mL)^{6}}{(mL)^{6}+6(mL)^{4}+45(mL)^{2}+225}
  \label{eqn20}\,.
 \end{eqnarray}
 for $n=1,3,5$.
 Note that the asymptotic form with respect to $mL$ in
 Eq.(\ref{eqn20}) is completely consistent with the case of $n$ even.
 Namely, the forms of the correction to the Newton's law are
 Eqs.(\ref{eqn18}) and (\ref{eqn19}) with regardless of $n$
 odd or even.
 Although the integral of Eq.(\ref{eqn14}) is performed by using the
 asymptotic form of Bessel functions, it is straightforward
 to calculate the integral Eq.(\ref{eqn14}) in the case of $n=1$.
 Consequently, the Newton's law on compactified
 $4$-brane embedded in six-dimensions with an infinite extra dimension
 is exactly given by
 \begin{eqnarray}
 V_{1}(r)&=&G_{N}\frac{M_{1}M_{2}}{r}+
  G_{N}LM_{1}M_{2}\int^{\infty}_{0}dm\;
  \frac{e^{-mr}}{r}\frac{2}{\pi}\frac{(mL)^2}{(mL)^2+1}\nonumber\\
  &=&G_{N}\frac{M_{1}M_{2}}{r}
  \left(1+\frac{2}{\pi}\frac{L}{r}
  \left\{\;1-\frac{r}{L}\left[\;\sin\left(\frac{r}{L}\right)
  {\rm ci}\left(\frac{r}{L}\right)-\cos\left(\frac{r}{L}\right)
  {\rm si}\left(\frac{r}{L}\right)\;\right]\;\right\}\right)\nonumber\\
 &&\label{eqn21}\,
 \end{eqnarray}
 where ${\rm ci}(x)=-\int^{\infty}_{x}\cos t/t\;dt$ and
 ${\rm si}(x)=-\int^{\infty}_{x}\sin t/t\;dt$ are cosine and sine
 integrals, respectively.
 Obviously, the behavior of Eq.(\ref{eqn21}) at long distance $r\gg L$
 is consistent with the case of $n=1$ in Eq.(\ref{eqn18}).
 On the other hand, at distance $r\ll L$, we can obtain the form up to 
 sub-leading order
 \begin{eqnarray}
  V_{1}(r)\sim G_{N}L\frac{M_{1}M_{2}}{r^{2}}
  \left(\;1-\frac{r^2}{L^{2}}\left(\gamma-1+\log\frac{r}{L}\right)\;\right)
  \,,\label{eqn22}
 \end{eqnarray}
 where $\gamma$ is the Euler constant.
 As indicated in Eq.(\ref{eqn18}), in this model at long distance
 the power of distance $r$ is a clue to searching for the invisible
 compactified radius with Planck length. 

 Since the possibility of theories with extra dimensions was indicated,
 the experiment of searching for the presence of extra dimensions are
 increasingly performed.
 From recent gravitational experiments,
 it is found that the gravitational force $1/r^{2}$ law is maintained up
 to  $0.218$ mm \cite{Hoyle:2000cv}.
 However it is unknown whether $1/r^{2}$ law is violated or not at about
 micrometer range.
 In the near future it is expected that the sophisticated equipment of 
 gravitational experiment will confirm the presence of extra
 dimensions. 
  
%
 \section{Summary}

 In the framework of a $(3+n)$-brane embedded in $(5+n)$ dimensions,
 where $n$ extra dimensions are compactified in Planck length and
 a dimension is infinite, we have shown that the wave function of gravity
 is described in terms of superposition of the Bessel functions of
 $(2+n/2)$-order. 
 We estimated the small correction to the four-dimensional
 Newton's law on the brane, and the correction term is $1/r^{n+2}$.
 In particular, for $n=1$ we can obtain the exact form of
 Newton's law.
 We presented a model indicated that a tiny remnant of higher dimensional
 world is observable in our world.

%
 

\begin{thebibliography}{99}
%
 \bibitem{Rubakov:1983bz}
 V.~A.~Rubakov and M.~E.~Shaposhnikov,
 ``Extra Space-Time Dimensions: Towards A Solution To The Cosmological
 Constant Problem,''
 Phys.\ Lett.\ B {\bf 125}, 139 (1983).
%
 \bibitem{Randall:1999ee}
 L.~Randall and R.~Sundrum,
 ``A large mass hierarchy from a small extra dimension,''
 Phys.\ Rev.\ Lett.\  {\bf 83}, 3370 (1999) [hep-ph/9905221].
%
 \bibitem{Randall:1999vf}
 L.~Randall and R.~Sundrum,
 ``An alternative to compactification,''
 Phys.\ Rev.\ Lett.\  {\bf 83}, 4690 (1999) [hep-th/9906064].
%
 \bibitem{Ito:2001jk}
 M.~Ito,
 ``Five-dimensional warped geometry with a bulk scalar field,''
 [hep-th/0109040].
%
 \bibitem{Ito:2001gk}
 M.~Ito,
 ``Various types of five dimensional warp factor and effective Planck  scale,''
 Phys.\ Lett.\ B {\bf 524}, 357 (2002) [hep-th/0109140].
%
 \bibitem{Lykken:1999nb}
 J.~Lykken and L.~Randall,
 ``The shape of gravity,'' JHEP {\bf 0006}, 014 (2000) [hep-th/9908076].
%
 \bibitem{Giddings:2000mu}
 S.~B.~Giddings, E.~Katz and L.~Randall,
 ``Linearized gravity in brane backgrounds,'' JHEP {\bf 0003}, 023 (2000)
 [hep-th/0002091].
%
 \bibitem{Karch:2000ct}
 A.~Karch and L.~Randall,
 ``Locally localized gravity,'' JHEP {\bf 0105}, 008 (2001)
 [hep-th/0011156].
%
 \bibitem{Csaki:2000fc}
 C.~Csaki, J.~Erlich, T.~J.~Hollowood and Y.~Shirman,
 ``Universal aspects of gravity localized on thick branes,''
 Nucl.\ Phys.\ B {\bf 581}, 309 (2000) [hep-th/0001033].
%
 \bibitem{Rubakov:2001kp}
 V.~A.~Rubakov,
 ``Large and infinite extra dimensions: An Introduction,'' [hep-ph/0104152].
%
 \bibitem{Dubovsky:2000av}
 S.~L.~Dubovsky, V.~A.~Rubakov and P.~G.~Tinyakov,
 ``Is the electric charge conserved in brane world?,''
 JHEP {\bf 0008}, 041 (2000) [hep-ph/0007179].
%
 \bibitem{Ito:2001fd}
 M.~Ito,
 ``Warped geometry in higher dimensions with an orbifold extra dimension,''
 Phys.\ Rev.\ D {\bf 64}, 124021 (2001) [hep-th/0105186].
%
 \bibitem{Hoyle:2000cv}
 C.~D.~Hoyle, U.~Schmidt, B.~R.~Heckel, E.~G.~Adelberger,
 J.~H.~Gundlach, D.~J.~Kapner and H.~E.~Swanson,
 ``Sub-millimeter tests of the gravitational inverse-square law: 
 A search  for 'large' extra dimensions,''
 Phys.\ Rev.\ Lett.\  {\bf 86}, 1418 (2001) [hep-ph/0011014].
 \end{thebibliography}
\end{document}